\documentclass[sigconf]{acmart}

\usepackage{booktabs} 
\usepackage{amsfonts}
\usepackage{amsthm}
\usepackage{url}
\usepackage{subcaption}
\usepackage{color}
\usepackage{graphicx}
\usepackage{pgfplots}
\usepackage[misc]{ifsym}

\usepackage{array}
\makeatletter
\newcommand{\thickhline}{%
    \noalign {\ifnum 0=`}\fi \hrule height 1pt
    \futurelet \reserved@a \@xhline
}
\newcolumntype{"}{@{\hskip\tabcolsep\vrule width 1pt\hskip\tabcolsep}}
\makeatother

\setcopyright{acmcopyright}

\begin{document}

\copyrightyear{2018}
\acmYear{2018}
\setcopyright{acmcopyright}
\acmConference[IDEAS 2018]{22nd International Database Engineering \& Applications Symposium}{June 18--20, 2018}{Villa San Giovanni, Italy}
\acmBooktitle{IDEAS 2018: 22nd International Database Engineering \& Applications Symposium, June 18--20, 2018, Villa San Giovanni, Italy}
\acmPrice{15.00}
\acmDOI{10.1145/3216122.3216126}
\acmISBN{978-1-4503-6527-7/18/06}

\title{Practical Study of Deterministic Regular Expressions from Large-scale XML and Schema Data}

\author{Yeting Li}
\affiliation{%
  \institution{State Key Laboratory of Computer Science, Institute of Software Chinese Academy of Sciences; University of Chinese Academy of Sciences}
}
\email{liyt@ios.ac.cn}

\author{Xinyu Chu}
\orcid{0000-0003-2300-8411}
\affiliation{%
  \institution{State Key Laboratory of Computer Science, Institute of Software Chinese Academy of Sciences; University of Chinese Academy of Sciences}
}
\email{chuxinyu17@mails.ucas.ac.cn}

\author{Xiaoying Mou}
\affiliation{%
  \institution{State Key Laboratory of Computer Science, Institute of Software Chinese Academy of Sciences; University of Chinese Academy of Sciences}
  }
\email{mouxy@ios.ac.cn}

\author{Chunmei Dong}
\affiliation{%
  \institution{State Key Laboratory of Computer Science, Institute of Software Chinese Academy of Sciences; University of Chinese Academy of Sciences}
  }
\email{dongcm@ios.ac.cn}

\author{Haiming Chen}
\authornote{Corresponding author.}
\affiliation{%
  \institution{State Key Laboratory of Computer Science, Institute of Software Chinese Academy of Sciences}
  \city{4 South Fourth Street, Zhong Guan Cun, Haidian District}
  \state{Beijing}
  \country{China}
  \postcode{100190}
  }
\email{chm@ios.ac.cn}

\renewcommand{\shortauthors}{Yeting et al.}

\begin{abstract}
Regular expressions are a fundamental concept in computer science and widely used in various applications. In this paper we focused on deterministic regular expressions (DREs). Considering that researchers didn't have large datasets as evidence before, we first harvested a large corpus of real data from the Web then conducted a practical study to investigate the usage of DREs. One feature of our work is that the data set is sufficiently large compared with previous work, which is obtained using several data collection strategies we proposed. The results show more than 98\% of expressions in Relax NG are DRE, and more than 56\% of expressions from RegExLib are DRE, while both Relax NG and RegExLib do not have the determinism constraint. These observations indicate that DREs are commonly used in practice. The results also show further study of subclasses of DREs is necessary. As far as we know, we are the first to analyze the determinism and the subclasses of DREs of Relax NG and RegExLib, and give these results. Furthermore, we give some discussions and applications of the data set. We obtain a DRE data set from the original data, which will be useful in practice and it has value in its own right. We find current research in new subclasses of DREs is insufficient, therefore it is necessary to do further study. We also analyze the referencing relationships among XSDs and define SchemaRank, which can be used in XML Schema design.
\end{abstract}

%
%


\maketitle

\section{Introduction}
Regular expressions (REs for short) is a fundamental concept in computer science and used in a large variety of applications, such as programming languages, search engines, text processing utilities, queries in graph databases and so on. The property and quality of RE and its subclasses have been studied extensively (See Related Work for details), but their practical study on large-scale actual data remains a challenge.

This paper focuses on deterministic REs (DREs for short). DREs are used in various applications such as the SPARQL query language for RDF \cite{Losemann2013The}, efficiently evaluating regular path queries \cite{Huang2014Answering} and AXML \cite{Abiteboul2005Regular}. Finding Extensible Markup Language (XML) use DREs extensively, we consider XML schema files as data source. XML is widely used for data exchange on the Web, which has been recommended as a standard for data exchange and data transmission by the World Wide Web Consortium (W3C) for data interchange and transmission. Structures of XML data are defined by schemas using schema languages. And the presence of a schema provides a lot of conveniences and advantages for various applications such as data processing, automatic data integration, static analysis of transformations and so on. Among the popular schema languages for XML, Document Type Definitions (DTDs) and XML Schema Definitions (XSDs) are recommended by the World Wide Web Consortium (W3C) \cite{Thompson2004XML,web1,web2}, and W3C specification requires that the content models of DTDs and XSDs must be DREs. Roughly speaking, determinism means when matching a word from left to right against an expression, a symbol can be matched to only one position in the expression without looking ahead. One immediate benefit of using DREs is efficient parsing. Indeed it gives a natural manner to define determinism in REs. As a result, several decision problems behave better for DREs than for general ones. For example, language inclusion for REs is PSPACE-complete but is tractable when the expressions are deterministic. It is known that DREs form a strict subclass of regular languages \cite{Br1998One}, which means that not every non-deterministic RE can be defined by a DRE. There have been a lot of research work related to DREs, see Related Work for some of them.

For practical studies of DREs, the real data set is the basis for relevant research work. However, this has been a weak point in the literature. For example, many researches on different subclasses of DREs were based just on hundreds of XSDs or DTDs. See Table~\ref{Datacomplete} for details. In this paper, we have obtained data from the Web, including RegExLib, Relax NG, XSD and DTD. RegExLib \cite{web4} is the main RE repository available on the Web, it contains multiple kinds of expressions for matching URIs, markup code, C style strings, pieces of Java code, SQL queries, spam, etc. Another popular schema language of XML, i. e., REgular LAnguage for XML Next Generation (Relax NG), is a standard of ISO (International Organization for Standardization) \cite{web3}. W3C specification requires that the content models of DTDs and XSDs must be DREs, while there is not a determinism restriction on Relax NG and RegExLib. Both XSD and Relax NG support interleaving operator, in which the interleaving supported by XSD is very limited and the interleaving supported by Relax NG is unlimited. And both XSD and RegExLib support counting operator. So it is representative to take them as examples to investigate the practical usage of DREs.

One feature of our work is that the data set is sufficiently large compared with previous work. Another point different with the precious is that we also gain expressions in a repository called RegExLib\footnote{\url{http://www.regexlib.com/}}. Harvesting a large corpus of DTDs, XSDs and Relax NGs from the Web is not an easy task, because of the two reasons: One is that although there are many schema files on the Web, they exist in different forms and locations so they can not be directly obtained in batches. Another one is the lack of standardized large-scale databases of schemas. Previous researchers gain schema files just from some local source, e.g. Bex et al. studied on 109 DTDs and 93 XSDs downloaded from the XML Cover pages repository in \cite{Bex2006Inference}. However, we have made good use of search engines and project hosting platforms, through these methods of obtaining data without source restriction, the data obtained are much larger and representative than those of previous researchers. Detailed as follows, we proposed four data collection strategies: comprehensively utilizing Google search engine, path-ascending crawling, downloading and analyzing the Web sites and finding potential data, to attain more schema files from the Web. Finally, we obtained 276,371 data files including 124,326 DTDs, 134,816 XSDs, 13,946 Relax NGs and 3,950 RegExLib expressions. Such a global scope schemas is significative in analyze DREs, because studying the practicability of DRE requires data to reflect practical application as far as possible, and large-scale random schemas are in line with this requirement.

Using the data set we conducted an extensive study to investigate the practical usage of DREs. We discovered that more than 98\% of REs in Relax NG are deterministic, which is quite surprising, because Relax NG
does not have the determinism constraint for its content models \cite{web3}. And more than 56\% of REs from RegExLib are deterministic. These seem to indicate that DREs are commonly used in practice. By analysing the data set, we find it is necessary to further study new practical subclasses of DREs and we have proposed some ones, such as the one in \cite{Peng2015Discovering}. We also analyze the referencing relationships among XSDs and define SchemaRank, which can be used in XML Schema design.

Since in practice most content models used in DTDs and XSDs essentially consist of restricted of subclasses of DREs, so for practical purpose many researches focus on the study of subclasses of DREs and their practical usage (e,g., \cite{Bex2004DTDs,Bex2005Expressiveness,Bex2006Inference,Martens2004Complexity,Li2016Practical}). We study the usages of various subclasses of DREs using the data set. Our experiments show the current research on subclasses of DREs is still in the initial stage, and further study is necessary.

The main contributions of this paper are as follows:

1. We harvested a large corpus of data from the Web,
including RegExLib, Relax NG, XSD and DTD. The data set is sufficiently large compared with previous work.

2. Using the data set we investigate the practical usage of DREs. The results seem to indicate that DREs are commonly used in practice. We also study the usages of various subclasses of DREs, our experiments show further study is necessary. As far as we know, we are the first to analyze the determinism of Relax NG and RegExLib and to analyze the subclasses of DREs of Relax NG and RegExLib, and give these results.

3. We construct a large-scale DRE data set from the data set, and use normalized DREs (see Section~\ref{normalization} for its definition) to get a compact and enhanced set. The obtained data set can be used as a basis for varies applications of DREs, and it has value in its own right.

4. We define SchemaRank to measuring reference relationships between XML schemas and the XML schema which is valid and important can use to schema design, (semi-)automatic repair and so on.

\textbf{Related Work.}

{\it Data collection and complexity analysis.} As shown in Table~\ref{Datacomplete}, Choi et al. \cite{Bray2004Extensible} provided statistics on some structures of real DTDs in 2002. Bex et al. \cite{Bex2004DTDs} studied the new features of XSDs and the complexity of REs in 2004, then they focused on expressiveness of XSDs \cite{Bex2005Expressiveness} in 2005. Barbosa et al. \cite{Barbosa2006Studying} researched the usage and complexity of DTDs in 2006. Laender et al. \cite{Laender2010An} analyzed which XSD constructs are more and less frequently used in 2008. Grijzenhout et al. \cite{Grijzenhout2011The} studied the the quality of the XML Web in 2011. Bj\"{o}rklund et al. \cite{Br1998One,Bj2015Efficient} focused on the usage of the counting operator in REs in 2015. Li et al. \cite{Li2016Practical} researched the complexity and usage subclasses of REs in DTDs and XSDs in 2016.

{\it Determinism.} To determine whether a standard RE is deterministic,
Br\"{u}ggemann-Klein \cite{Ggemann1992Regular} gave an $O(|\Sigma_E||E|)$ time algorithm, where $|\Sigma_E|$ is the set of distinct symbols in $E$. For REs with counting, Kilpel\"{a}inen \cite{Kilpel2011Checking} presented an $O(|\Sigma_E||E|)$ time algorithm by examining the marked expression. Chen and Lu \cite{Chen2011Assisting} investigated algorithms that check the determinism of the input standard RE and provide diagnostic information if the expression is not deterministic, they also gave an $O(|\Sigma_E||E|)$ time algorithm for checking determinism of REs with counting \cite{Chen2015Checking}. Peng et al. \cite{Peng2015Deterministic} proposed an $O(|\Sigma_E||E|)$ time algorithm for checking determinism of REs with interleaving. Groz and Maneth \cite{Groz2017Efficient} put forward the first $O(|E|)$ time algorithm for checking determinism of standard REs and REs with counting.

\section{Preliminaries}
\subsection{Regular Expressions With Counting and Interleaving}
Let $\Sigma$ be an alphabet of symbols. Sometimes symbols are
also called elements. The set of all finite words over $\Sigma$ is
denoted by $\Sigma^*$. The empty word is denoted by $\varepsilon$. A standard
RE over $\Sigma$ is defined as: $\phi$, $\varepsilon$
or $a \in \Sigma$ is a RE, the union $E_1|E_2$, the
concatenation $E_1E_2$, or the Kleene-star $E^*_1$ is a RE for REs $E_1$ and $E_2$. Let $\mathbb{N}$ denote the set $\{0, 1, 2,...\}$. A RE with counting
and interleaving is extended from standard REs by further using the
numerical iteration operator $E^{[m,n]}$ and the interleaving operator $E_1\&E_2$. The bounds $m$ and $n$ satisfy the following conditions: $m\in\mathbb{N}$, $n\in\mathbb{N}\backslash\{0\}\cup\{\infty\}$, and $m \le n$. We use $s_1\&s_2$ to denote the set of strings obtained by $s_1$ and $s_2$ in every possible way. For $s_1,s_2\in \Sigma^*$ and $a,b\in\Sigma$, $s_1\&\varepsilon = \varepsilon\&s_1 = \{s_1\}$ and
$as_1\& bs_2 = \{a(s_1\& bs_2)\}\cup \{b(as_1\& s_2)\}$.
\subsection{Deterministic Regular Expressions}
For a RE we can mark symbols with subscripts so that in the marked expression each marked symbol occurs only once. E.g. $(a_1+b_1)^*a_2$ is a marking of the expression $(a+b)^*a$. The marking of $E$ is denoted by $\overline{E}$. The same notation will also be used for dropping subscripts from the marked symbols: $\overline{\overline{E}}=E$. We extend the notation for words and sets of symbols in the obvious way. It will be clear from the context whether $\overline{\cdot}$ adds or drops subscripts. An expression $E$ is deterministic if and only if for all words $uxv$, $uyw$ $\in$ $L(\overline{E})$ where $|x|=|y|=1$, if $x\neq y$ then $\overline{x}\neq\overline{y}$.

Deterministic requires that the matching position is unique when matching sentences to the regular expression. For example, $a(a)^*$ is deterministic while $(a)^*a$ is not, although the languages they define are equivalent. For $a_2$ $\in$ $L((a)^*a)$ and $a_1a_2$ $\in$ $L((a)^*a)$, set $u=\varepsilon$, $x=a$, $y=a$, $v=\varepsilon$, $w=a$, we can see $x=y=a$ but $\overline{x}(a_2)\neq \overline{y}(a_1)$, so $(a)^*a$ is not a deterministic regular expression.

\subsection{Definitions}
There follows some subclasses of deterministic regular expressions which are used frequently:

\begin{definition}[SORE] Let $\Sigma$ be an alphabet. A single-occurrence RE is a standard RE over $\Sigma$ in which every terminal symbol occurs at most once. E.g. $(a^*b^{[0,2]})^+$ is a SORE but $(a^*b^{[0,2]}a^*)^+$ is not, although the languages they define are equivalent.
\end{definition}

\begin{definition}[Simplified CHARE] A Simplified CHARE (also called CHARE) is a SORE over $\Sigma$ of the form $f_1 ... f_n$ where $n\ge 1$. Factor $f_i$ is an expression of the form $(a_1 +...+ a_m), (a_1 +...+ a_m)^?, (a_1 +...+ a_m)^*,(a_1 +...+ a_m)^+$ where $m\ge 1$ and $a_i \in \Sigma$. E.g. $a(b|c)^*d^+(e|f)^?$ is a Simplified CHARE, but $(ab|c)^*$ is not because $ab$ is a non-terminal, $(a^*|b^?)^*$ is also does not belong to Simplified CHARE because $a^*$ or $b^?$ contains unary operator.
\end{definition}

\begin{definition}[eSimplified CHARE] An eSimplified CHARE is a SORE over $\Sigma$ of the form $f_1 ... f_n$ where $n\ge 1$. Factor $f_i$ is an expression of the form $(b_1 +...+ b_m), (b_1 +...+ b_m)^?, (b_1 +...+ b_m)^*,(b_1 +...+ b_m)^+$ where $m\ge 1$ and $b_i$ is $a_i$ or $a_i^+$ where $a_i \in \Sigma$. E.g. $a|b^+$ is a eSimplified CHARE, but $(ab|c)^*$ is not because $ab$ is a non-terminal.
\end{definition}

\section{A Practical Study}
We conducted an extensive study to investigate the practical
usage of DREs. We investigated expressions in the RegExLib
repository, and carried out a thorough search for DREs
in XML schemas on the Web.
\subsection{Data Set}
We have obtained data from the Web, including RegExLib, DTDs, XSDs and Relax NGs. As we have mentioned, it is representative to take them as examples to investigate the practical usage of DREs (with counting and interleaving).
\subsubsection{RegExLib data}

The RegExLib library describes itself as the Internet's first RE Library which supports RE with counting. It contains multiple kinds of expressions to match URIs, markup code, C style strings, pieces of Java code, SQL queries, spam, etc. We crawled the expressions of the library and obtained 3,950 expressions after removing duplicates.
\subsubsection{Harvesting Schema files from the Web}

The strategies we proposed for obtaining schema files from the Web are briefly explained in the following.

\textbf{Comprehensively utilizing Google search engine.} A
large number of XML schema files on the Web are needed
to find URLs through search engines. We use the API
of Google's Custom Search Engine to get resources. We
use filetype and site instructions to obtain the URLs of
DTDs, XSDs and Relax NGs.

During the experiment, we save the URLs of XSDs,
DTDs, Relax NGs to there corresponding documents,
and remove duplicate URLs at the same time. When downloading a
schema file, we build the folders layer by layer according to its URL directory, and then store it locally. The advantage of doing so is that when an XML schema file
is in mistake or ambiguity, we can trace back to its URL
on the Web according to the directory stored and check it.
These methods are also used in the following three strategies.

\textbf{Path-ascending crawling.} We use an
example to show the use of path-ascending
crawling strategy. When given a seed URL
of \url{http://52north.org/schema/users/1.0/users.xsd},
it will attempt to crawl \url{52north.org},
\url{52north.org/schema}, \url{52north.org/schema/users}, and
\url{52north.org/schema/users/1.0}. We find that the path-ascending strategy is very effective in finding isolated
resources, or resources for which no link could be found in regular crawling.

\textbf{Downloading and analyzing the Web sites.} Some
URLs of schema files come from Maven\footnote{\url{http://repo1.maven.org}}, it turns out
that DTDs, XSDs, Relax NGs hidden in the JARs/ZIPs
of Maven cannot be found through the search engines, so
we downloaded all files with suffixes of jar or zip and
then extract them locally. We then filter out files of other
types, keeping only XSDs, DTDs and Relax NGs. We treat GitHub projects similarly.

\textbf{Finding potential data.} A site may have different formats of XML schema files, that is, a site with Relax NGs are also likely to have XSDs and DTDs, so
we also conduct a crossed search. This method is proved
to be effective in our experiments and give an access to
a large number of URLs of schema files.

Finally, we collected 273,088 schemas files in total, the amount of which is 36 times of Grijzenhout et al. \cite{Grijzenhout2011The}, 34 times of Bj\"{o}rklund et al. \cite{Bj2015Efficient} and 37 times of Li et al. \cite{Li2016Practical}. In details, DTD files are expanded by 39 times compared with
Li et al. \cite{Li2016Practical} and XSD files expanded by 15 times
compared Bj\"{o}rklund et al. \cite{Bj2015Efficient} and Relax NGs expanded to 41 times of Grijzenhout et al. \cite{Grijzenhout2011The}. The
details are shown in Table~\ref{Datacomplete}. We use RNG to represent Relax
NG in this and following tables. Our repository can be found at \url{https://github.com/clRE/XMLSchemas}.

\begin{table}
\centering
\vspace{0.01cm}
\setlength{\abovecaptionskip}{0cm}
\setlength{\belowcaptionskip}{0cm}
  \caption{Data obtained over the years}
  \label{Datacomplete}
  \begin{tabular}{|c|c|c|c|c|c|}
  \hline
    Year & DTDs & XSDs & RNGs & Total & Work \\\hline
     2002 & 60 & N/A & N/A & 60 & Choi \\
          &     &     &    &    & et al. \cite{Choi2002What} \\\hline
     2004 & 109 & 93 & N/A & 202 & Bex et al. \cite{Bex2004DTDs}\\\hline
     2005 & N/A & 819 & N/A & 819 & Bex et al. \cite{Bex2005Expressiveness} \\\hline
     2005 & N/A & 199 & N/A & 199 & Laender \\
          &     &     &     &     & et al. \cite{Laender2010An}\\\hline
     2006 & 75 & N/A & N/A & 75 & Barbosa\\
          &     &     &     &     & et al. \cite{Barbosa2006Studying}\\\hline
     2007 & N/A & 697 & N/A & 607 & Bex et al. \cite{Bex2007Inferring}\\\hline
     2008 & N/A & 223 & N/A & 223 & Laender \\
          &     &     &     &     & et al. \cite{Laender2010An}\\\hline
     2011 & 3,087 & 4,141 & 337 & 7,565 & Grijzenhout \\
          &      &      &     &      & et al. \cite{Grijzenhout2011The}\\\hline
     2015 & N/A & 8,000+ & N/A & 8,000+ & Bj\"{o}rklund\\
          &     &       &     &       & et al. \cite{Bj2015Efficient} \\\hline
     2016 & 2,427 & 4,859 & N/A & 7,286 & Li et al. \cite{Li2016Practical}\\\hline
     2018 & 124,326 & 134,816 & 13,946 & 273,088 & Ours \\\hline
\end{tabular}
\end{table}
\subsection{Data Preprocessing}

On the data collected, as for REs in RegExLib, we remove duplicates and then transform to our syntax; as for schema files, we preprocess with the following steps.

(1). Schemas normalization. We remove annotations, redundant blank lines and white spaces
to get pretty and normalized files. (2). Duplicate file removal. Then
we use the algorithm SimHash \cite{Pi2009SimHashbasedEA} to check if two files are
the same and delete the duplicate ones. (3). Well-formedness
and validity checking. Using the XMLSpy tool, we check the
well-formedness and validity of our schema files, and remove
the bad ones, which ensures the accuracy of our research.
(4). Schema2re. We parse content models of DTD, XSD and
Relax NG files into corresponding REs.

At last, we obtain files: 29,414 DTDs, 38,554 XSDs, 4,526
Relax NGs and extract 118,242, 476,804 and 509,267 regular
expressions from them respectively.
\subsection{Data Analysis}
\subsubsection{Determinism}

XSD supports a very limited form of interleaving while Relax NG supports unlimited interleaving. To analyze determinism of Relax NG, we need tools that can decide determinism of unlimited interleaving. Fortunately, we have solved this problem and have tools for deciding determinism of RE with counting and unlimited interleaving. So we can analyze determinism of any RE with counting and interleaving, including Relax NG. This forms the basis of the present experiments.

In concrete, we studied the determinism of REs generated by schemas and the REs from RegExLib respectively. The results are shown in Figure~\ref{fig:Deterministic}, containing the percent of DREs in whole expressions and DREs of two non-SORE subclasses in all REs. Note that the subclass CHARE accounts for 78.96\% of all REs, eCHARE accounts for 79.19\%. As mentioned, determinism is required by W3C specification for content models of DTDs and XSDs. On the contrary, Relax NG does not require patterns to be ``deterministic" or ``unambiguous", so does the expressions from RegExLib. We discovered that more than 98\% of expressions in Relax NG are deterministic and more than 56\% for RegExLib. These observations indicate that DREs are commonly used in practice. Nondeterministic REs in DTDs or XSDs shows the DTDs or XSDs do not satisfy the specifications.
\definecolor{blue1}{RGB}{145, 225, 255}
\definecolor{blue2}{RGB}{115, 178, 222}
\definecolor{blue3}{RGB}{95, 145, 185}
\definecolor{blue4}{RGB}{65, 105, 125}
\begin{figure}
\centering
\vspace{0cm}
\setlength{\belowcaptionskip}{0cm}
\centering
\begin{tikzpicture}
\begin{axis}[
    xticklabels={\small{DTD},\small{XSD},\small{RNG},\small{RegExLib}},
    xtick={0,1,2,3},
    x tick label style={rotate=0,xshift=3,yshift=3},
	ylabel={\small{DREs in all REs$(\%)$}},
    y label style={rotate=0,xshift=0,yshift=-10},
	enlargelimits=0.01,
    ymajorgrids=true,
    legend pos=south east,
	legend style={at={(1.35,0)},},
	ybar interval=1,
    xmajorgrids=false,
    width=190,
    height=160,
]
\addplot[]
	coordinates {(0,0)(1,0)(2,0)(3,0)(4,0)};
\addplot[draw=black,fill=blue1]
	coordinates {(0,0.9883)(1,0.9993)(2,0.9825)(3,0.5656)(4,0)};
\addplot[draw=black,fill=blue2]
	coordinates {(0,0.8761)(1,0.9157)(2,0.6098)(3,0.1806)(4,0)};
\addplot[draw=black,fill=blue4]
	coordinates {(0,0.8798)(1,0.9184)(2,0.6106)(3,0.2143)(4,0)};
\addplot[]
	coordinates {(0,0)(1,0)(2,0)(3,0)(4,0)};
\legend{,\tiny{Whole},\tiny{CHARE},\tiny{eCHARE}}
\end{axis}
\end{tikzpicture}
\caption{Determinacy of REs}
\label{fig:Deterministic}
\end{figure}
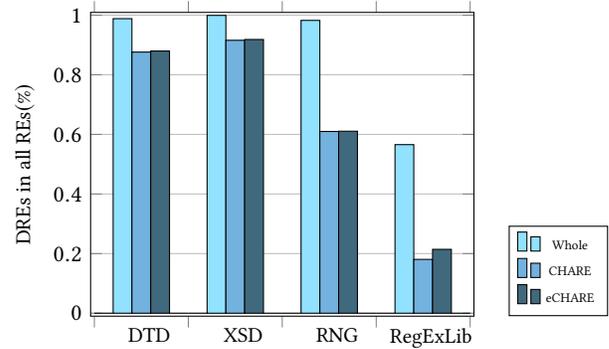

\subsubsection{Usage of subclasses of DREs}

In practice many DTDs and XSDs contains subclasses of DREs so we study the usages of various subclasses of DREs. We use the definitions of the subclasses of restricted REs from \cite{Li2016Practical}.
Existing subclasses of DREs are all defined on standard REs, i,e., SORE \cite{Bex2006Inference}, Simplified CHARE \cite{Bex2006Inference}, eSimplified CHARE \cite{Li2016Practical}. And the last two are subclasses of SORE. In the experiments we also include three new subclasses: SOREwCorI, SOREwC, SOREwI, stands for SORE with counting and interleaving, with counting and with interleaving, respectively. The results are presented in Table~\ref{Subclasses}, and for the first three subclasses reasons dissatisfying the corresponding definitions are also given. The results show the percentage of subclasses of DREs in RegExLib are not high,
while a large percentage of DREs in DTDs, XSDs and Relax
NGs belongs to SOREwCorI. Note Bex et al. \cite{Bex2004DTDs} analyzed
SORE with 109 DTDs and 93 XSDs, and Bj\"{o}rklund et al. \cite{Bj2015Efficient}
analyzed SORE with more than 8,000 XSDs, the results for
SORE are similar.

Our experiments show the current research on subclasses of DREs have the following weak points: no experiments on large-scale real data; no subclass of DREs that belongs to non-SORE; although in experiments SORE and its subclasses take a high percentage, actually they are overly used in practice; and existing subclasses of DREs all defined on standard REs, lacking counting and/or interleaving. Our conclusion is that the current research on subclasses of DREs is still in the initial stage, and further studies are necessary. See Section~\ref{sec4} for detailed discussions.
\begin{table}
\centering
\vspace{0cm}
\setlength{\abovecaptionskip}{0cm}
\setlength{\belowcaptionskip}{0cm}
  \caption{The proportions of subclasses of DREs}
  \label{Subclasses}
  \begin{tabular}{|c|c|c|c|c|}
  \hline
    Subclasses & DTDs & XSDs & RNGs & RegEx\\
    &(\%)&(\%)&(\%)&Lib(\%)\\\hline
    \hline
    SOREs &93.43& 94.12& 71.16& 15.04\\\hline
    nonstandard expression& 0.00& 5.47& 25.03& 52.13\\\hline
    2-OREs& 4.98& 0.33& 2.44& 10.70\\\hline
    3-OREs& 0.43& 0.04& 0.55& 6.20\\\hline
    4-OREs& 0.32& 0.02& 0.28& 5.17\\\hline
    5-OREs& 0.66& 0.01& 0.37& 1.94\\\hline
    6-OREs& 0.08& 0.01& 0.02& 2.35\\\hline
    more than 7& 0.08& 0.00& 0.15& 4.24\\\hline
    \hline
    Simplified CHAREs& 85.33& 90.58& 59.40& 12.54\\\hline
    not a SORE &6.57& 0.41& 3.81& 32.83\\\hline
    not a terminal symbol &6.24 &1.87 &7.98 &2.46\\\hline
    occur unary operators &1.86 &1.67& 3.79 &0.05\\\hline
    nonstandard expression &0.00 &5.47 &25.03 &52.13\\\hline
    \hline
    eSimplified CHAREs& 86.18& 90.92& 60.98& 12.59\\\hline
    not a SORE &6.57& 0.41& 3.81& 32.83\\\hline
    not a terminal symbol& 6.24& 1.87& 7.98& 2.46\\\hline
    the unary operator *or?& 1.00& 1.34& 2.20& 0.00\\\hline
    nonstandard expression& 0.00& 5.74& 25.03& 52.13\\\hline
    \hline
    SOREwC& 93.43& 96.88& 71.16& 21.84\\\hline
    \hline
    SOREwI& 93.43 &96.82& 96.05& 15.04\\\hline
    \hline
    SOREwCorI& 93.43& 99.57& 96.05& 21.84\\\hline
\end{tabular}
\end{table}

\subsubsection{Complexity}
\textbf{Star height \cite{Bala2002Intersection}.}
The star height of a RE $E$ over the alphabet $\Sigma$, denoted by $h(E)$, is a nonnegative integer defined recursively as follows:

1. $h(E)=0$, if $E=\phi$ or $a$ for $a \in \Sigma$;

2. $h(E)$ = $max\{h(E_1), h(E_2)\}$, if $E=(E_1+E_2)$ or $E=(E_1・E_2)$ where $E_1$, $E_2$ are REs over $\Sigma$;

3. $h(E)=h(E_1)+1$, if $E=(E_1)^*$ and $E_1$ is a RE over $\Sigma$.

The results are shown in Table~\ref{starheight}.

\textbf{Nesting Depth.}
The nesting depth of a RE $E$ over the alphabet $\Sigma$, denoted by $ND(E)$, is a nonnegative integer defined recursively as follows:

1. $ND(E)=0$, if $E=\phi$ or $a$ for $a \in \Sigma$;

2. $ND(E)$ = $max\{ND(E_1), ND(E_2)\}$, if $E=(E_1+E_2)$, $E=(E_1\&E_2)$ or $E=(E_1・E_2)$ where $E_1$, $E_2$ are REs over $\Sigma$;

3. $ND(E)=ND(E_1)+1$, if $E=(E_1)^*$, $E=(E_1)^+$, $E=(E_1)^?$ or $E=(E_1)^{[m,n]}$ for $E_1$ is a RE over $\Sigma$.

The results are shown in Table~\ref{nestingdepth}.

\begin{figure}
  \includegraphics[scale=0.51]{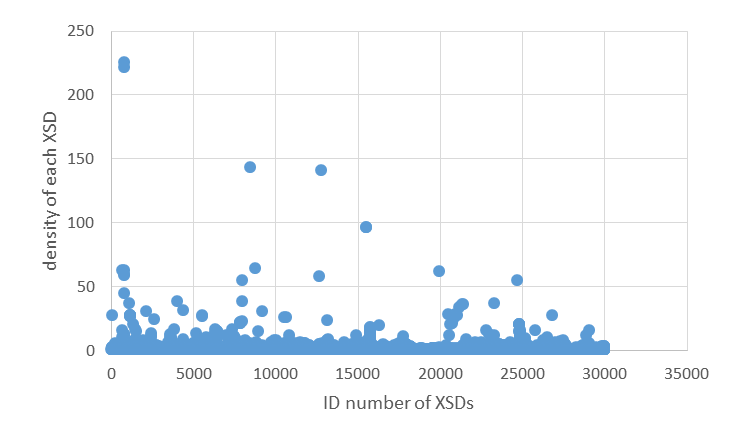}
  \includegraphics[scale=0.51]{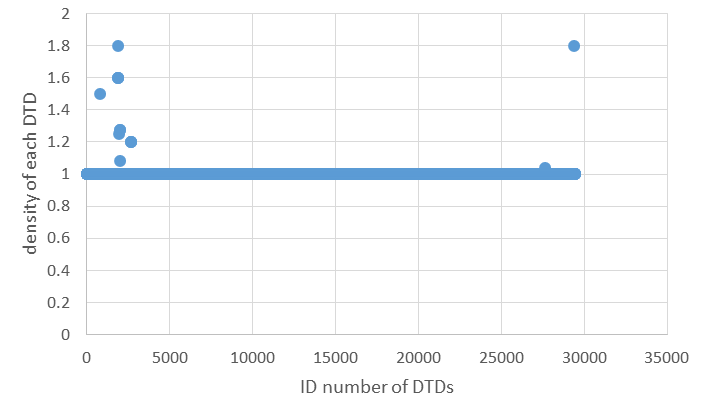}
  \includegraphics[scale=0.5]{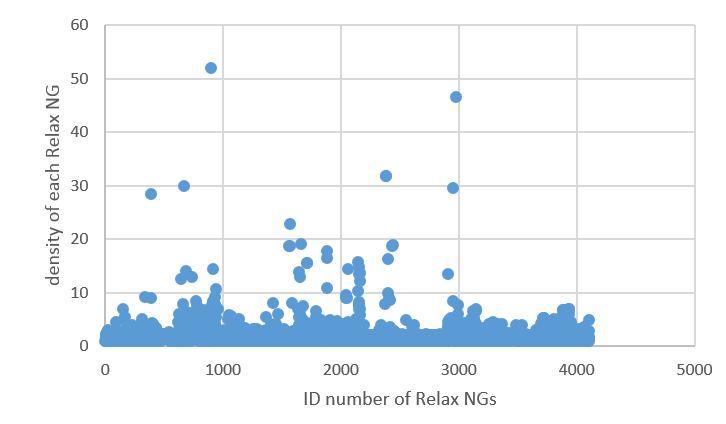}
\caption{Density of XSDs, DTDs and Relax NGs}
\label{fig:Density}
\end{figure}

\textbf{Density \cite{Nordmann2011Algorithmic}.}
The density of a schema is defined as the number of elements
occurring in the right hand side of its rules divided by the number
of elements. The formula:
$$d=\frac{1}{N}
\sum_{i=1}^N|A_i|$$
, where $N$ is the total number of element definitions occurring in
this schema, $A_i$ is the string in the right hand of a rule, and $|A_i|$
denotes the size of $A_i$. The XML Schema files with bigger density
value have higher complexity.
Experiment shows that the average density of Relax NGs, XSDs and DTDs are 1.8689, 1.3476 and 1.0002, respectively. We made Figure~\ref{fig:Density} to show the density of the three kinds of XML schemas. 
\begin{table}
\centering
  \caption{Star height in DTDs, XSDs and RNGs}
  \label{starheight}
  \begin{tabular}{|c|c|c|c|}
  \hline
    Star height &DTDs(\%)&XSDs(\%)&RNGs(\%)\\\hline
    0& 26.56& 65.27& 66.42\\\hline
    1& 71.72& 34.11& 32.86\\\hline
    2& 1.69& 0.57& 0.72\\\hline
    3& 0.03& 0.05& 0\\\hline
\end{tabular}
\end{table}
\begin{table}
\centering
  \caption{Nesting depth in DTDs, XSDs and RNGs}
  \label{nestingdepth}
  \begin{tabular}{|c|c|c|c|}
  \hline
    Nesting depth&DTDs(\%)&XSDs(\%)&RNGs(\%)\\\hline
    0& 94.58&99.24& 91.31\\\hline
    1& 4.60& 0.73& 8.45\\\hline
    2& 0.58& 0.02&0.09\\\hline
    3& 0.24& 0.01& 0.15\\\hline
\end{tabular}
\end{table}

\subsubsection{Counting and interleaving}

We analyze counting and interleaving in REs. Note that we only analyze non-trivial counters as \cite{Bj2015Efficient}, which means in the form $r^{m,n}$ one of $m$ or $n$ is at least two. The result is: 2.76\% of expressions from XSD, and 52.25\% of expressions from RegExLib has counting. 2.71\% of expressions from XSD, and surprisingly 25.03\% of expressions from Relax NG has interleaving. We further analyze the interleaving of expressions from Relax NG. In the 4500 files we obtained, the feature $\langle interleave\rangle$ appears 55898 times in total. This shows RE with interleaving are quite often in Relax NG, so extending RE with interleaving is useful and it is necessary to research the application of RE with interleaving in Relax NG.
\section{Discussions and Applications\label{sec4}}
\subsection{SchemaRank}
\subsubsection{Definition and results on XSDs}
\begin{figure}
\centering
\includegraphics[height=4.5cm, width=6.5cm]{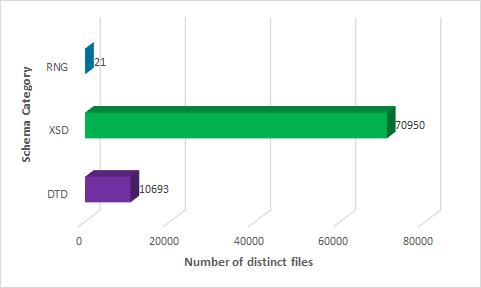}
\caption{Example of SchemaRank relations}
\label{fig:xs}
\end{figure}

Considering XSD accounts for the highest proportion of the XML Schemas obtained, and what's more, our experiment shows XSD is the most widely used schema when defining the XML (Figure~\ref{fig:xs}). To gain the result, we collected 81664 XML files using the same strategies as harvesting Schema files from the Web, then focused on the Schemas which these XML files are based on. The graph shows that XSD has an absolute advantage in defining XML, at least nowadays.

Based on above-mentioned reasons, we concentrate our attention on the importance of Schema files
by analyzing the referencing relationships among XSD files.
We construct the Schema files network diagram, where each
Schema file is regarded as a node, an $\langle import\rangle$ statement
denotes an oriented edge from the analysed Schema to the
referenced Schema. The reference digraph can help us to intuitively understand the organization of XML Schemas. The PageRank \cite{web5} algorithms is known
to evaluate importance of nodes, so we utilize it to calculate
the importance of Schema files. For each XSD file, the initial
Rank value is $1/n$ ($n$ is the total number of files). Then
we use PageRank algorithm to obtain the final rank value
corresponding to the importance of each node, which is defined as SchemaRank. $SR(p)=PR(p)*times(p)$, the $times(p)$ is the schema occurred times in XSD files.

\begin{figure}
\centering
\includegraphics[height=3.8cm, width=7.2cm]{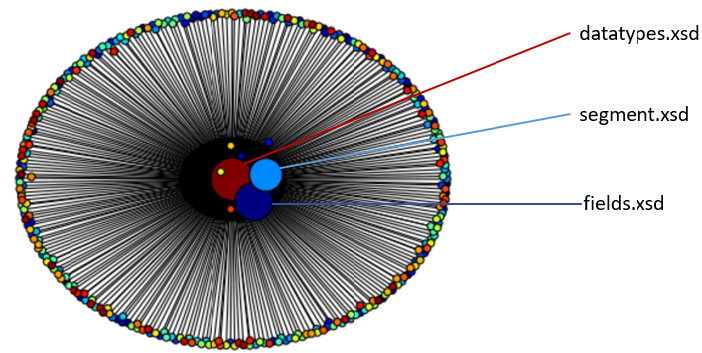}
\caption{Example of SchemaRank relations}
\label{fig:sr}
\end{figure}

Figure~\ref{fig:sr} shows the local relations between the nodes with top three SchemaRank values and other nodes which have direct references to them. Table~\ref{tab:sr} were made to show XSDs in our large-scale schema data who have the highest SchemaRank values. After analyzing the source of the top XSDs, we get the following results: most of the XSDs with high SchemaRank values are obtained from GitHub, which precisely confirms the correctness and efficiency of our data obtain strategy -- Downloading and analyzing the Web sites. The second source is obtained through Google, on www.w3c.org web site and www.opencms.org web site (a professional open source web content management system).

\subsubsection{Experiments on a subset of XSDs}
\begin{figure}
  \includegraphics[scale=0.45]{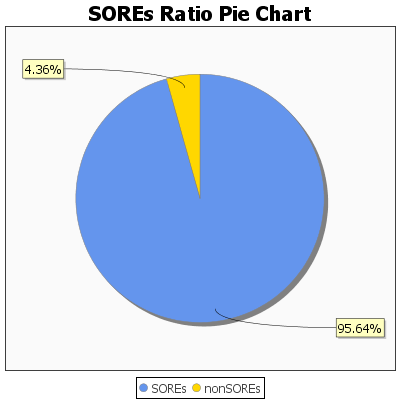}
  \includegraphics[scale=0.45]{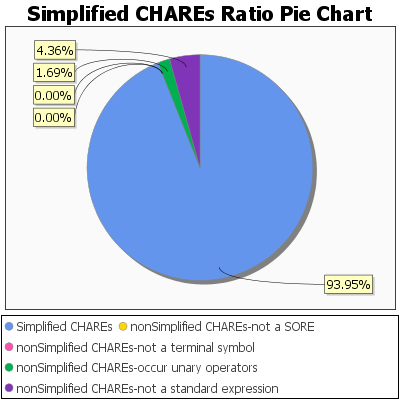}
  \includegraphics[scale=0.45]{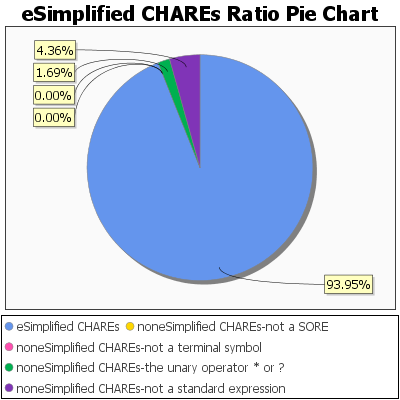}
\caption{The proportions of subclasses of DREs parsed from a subset of XSDs}
\label{fig:importsubclass}
\end{figure}

After constructing the schemas network diagram, we selected 42 important schema files who have obviously higher SchemaRank values than other schemas and focused on the 39,880 regular expressions parsed from these schema files. Satisfactorily, these expressions are all deterministic regular expressions, which further confirms the wide practicability of DRE. We also analyzed which subclasses these expressions belong to, mainly studied the proportion of the three subclasses, SORE, Simplified CHARE and eSimplified CHARE, shown in Figure~\ref{fig:importsubclass}. The results show that it is still necessary to define a new more practical subclass.

\begin{table*}
  \centering
  \caption{Top 15 XSDs ordered by SchemaRank value}\label{tab:sr}
  \begin{tabular}{c|c}
     \thickhline
     \small{XML Schemas Definition} & \small{SchemaRank} \\\thickhline
     \scriptsize{https://github.com/kdar/health/blob/master/hl7x/gen/vendor/2.5/segments.xsd} &\footnotesize{0.021643} \\\hline
     \scriptsize{https://github.com/kdar/health/blob/master/hl7x/gen/vendor/2.7/fields.xsd} & \footnotesize{0.017222} \\\hline
     \scriptsize{https://github.com/cqframework/healthedecisions/blob/master/specification/src/schema/common/datatypes.xsd} & \footnotesize{0.014761} \\\hline
     \scriptsize{http://user47094.vs.easily.co.uk/netex/schema/1.00/xsd/netex\_framework/netex\_responsibility/netex\_relationship-v1.0.xsd} & \footnotesize{0.009911} \\\hline
     \scriptsize{https://github.com/ot4i/dfdl-hl7-tutorial/blob/master/src/HL7-2.7/Z\_Segments.xsd} & \footnotesize{0.009299} \\\hline
     \scriptsize{https://github.com/robert197/exchangerxml/blob/master/samples/open\%20applications/OAGIS/Resources/Fields.xsd} & \footnotesize{0.008753} \\\hline
     \scriptsize{http://user47094.vs.easily.co.uk/netex/schema/1.03/xsd/netex\_framework/netex\_responsibility/netex\_relationship\_support-v1.0.xsd} & \footnotesize{0.008717} \\\hline
     \scriptsize{http://user47094.vs.easily.co.uk/netex/schema/1.03/xsd/netex\_framework/netex\_responsibility/netex\_version\_support-v1.0.xsd} & \footnotesize{0.007655} \\\hline
     \scriptsize{http://www.omg.org/spec/CDSS/20101201/voc.xsd} &\footnotesize{ 0.005981 } \\\hline
     \scriptsize{https://github.com/dmichael/amazon-mws/blob/master/examples/xsd/amzn-base.xsd} &\footnotesize{ 0.005800} \\\hline
     \scriptsize{https://www.retsinformation.dk/offentlig/xml/schemas/2015/06/22/Meta.LexDania\_2.1.xsd} &\footnotesize{ 0.005648} \\\hline
     \scriptsize{https://tools.oasis-open.org/version-control/browse/wsvn/sca-assembly/SCA\_XSDS/sca-core-1.1-cd06.xsd} & \footnotesize{0.005302} \\\hline
     \scriptsize{http://w3.energistics.org/energyML/data/common/v2.1/xsd\_schemas/gml/3.2.1/gml.xsd} &\footnotesize{ 0.004891} \\\hline
     \scriptsize{http://www.cafeconleche.org/books/bible3/source/25/source.xsd} &\footnotesize{ 0.004559} \\\hline
     \scriptsize{http://w3.energistics.org/schema/gml\_resqml\_v1.0\_profile/gml/3.2.1/gml.xsd} &\footnotesize{ 0.004891} \\
     \thickhline
   \end{tabular}
\end{table*}

\subsubsection{Analysis and application}
In practice, there are often cases of XML Schema not well-formed or invalid. According to \cite{Grijzenhout2013The}, only 24.8\% of the XML on the Web contain a reference to a DTD or XSD, of which just one third is actually valid. Automatic Schema repair seems promising and the SchemaRank can make a contribution when there is an element or attribute used that is not defined in the schema. Another application is XML Schema design \cite{Algergawy2010Element}, the SchemaRank can be used in the step of Schema complement, i.e., supplement type attribute to Schema fragment, which can make schemas more exact.

\subsection{The Data Set of DREs}
\subsubsection{DRE normalization\label{normalization}}

We normalize the DREs in the data set. A DRE is normalized if the symbols in the DRE, in the order from left
to right, are uniformly substituted by symbols $a_1, a_2, a_3$, ....
Note a repeatedly occurred symbol will be substituted by a
same symbol. An example is given in the following, suppose we have three original DREs: (1) $red,green,blue$; (2) $name,age,sex$; (3) $red,green,green$.  The normalized DREs are as follows: (1) $a_1, a_2, a_3$; (2) $a_1, a_2, a_3$; (3) $a_1, a_2, a_2$.

As a result, expressions with same or similar structures can
be merged, thus we get a more compact set. Moreover, the
normalized DREs cover more DREs than the original data set
(actually any DRE that has the same structure as (a subtree of)
a normalized DRE but is not contained in the original data set
will be covered).
\subsubsection{The construction of the DRE set}
We get DREs from the original data set and then get the normalized DREs from the
set of DREs, in which the number of expressions in the normalized DREs is only about 2.87\% of the set of DREs. Shown in Table~\ref{numberOf}. This normalized DRE data set will be valuable in applications and has value in its own right.
\begin{table}
\centering
\vspace{0cm}
\setlength{\abovecaptionskip}{0.1cm}
  \caption{Number of DREs}
  \label{numberOf}
  \begin{tabular}{|c|c|c|}
  \hline
    Type &Original DRE set &Normalized DRE set \\\hline
    DTD& 87,176& 3,767\\\hline
    XSD& 266,100& 14,771\\\hline
    RNG& 353,926& 2,791\\\hline
    RegExLib& 2,234& 724\\\hline
    Total& 709,436& 20,339\\\hline
\end{tabular}
\end{table}
\subsection{New Practical Subclasses}
Although the results in Table~\ref{Subclasses} show SORE has a high percentage, actually in SORE and its subclasses, overgeneralization is quite common as shown by the following example \cite{Boneva2015Schemas}. For instance, the DTD below corresponds to the one used in practice for the DBLP repository:

$dblp \to (article | book)^*$

$article \to (title | year | author)^*$

$book \to (title | year | author | editor | publisher)^*$

In this example, we shall require every $article$ to have exactly one $title$, one $year$, and one or more $author$. A $book$ may additionally contain one $publisher$ and may also have one or more $editor$ instead of $author$. But this DTD allows an article to contain any number of $title$, $year$, and $author$ elements. A book may also have any number of $title$, $year$, $author$, $editor$, and $publisher$ elements. These REs are clearly overgeneralization because they allow documents that do not follow the intuitive guidelines set out earlier e.g., a document containing an $article$ with two $title$'s and no $author$ should not be admissible.

In fact, the following schema captures precisely the intuitive requirements for the DBLP repository:

$dblp \to article^* \& book^*$

$article \to title \& year \& author^+$

$book \to title \& year \& publisher^? \& (author^+ | editor^+)$

This suggests that we can get more accurate result by using subclasses with counting and interleaving. Actually there have been some preliminary work, e.g., \cite{Boneva2015Schemas,Peng2015Discovering,Zhang2018Inference}.

This shows the necessity to further study new practical subclasses, based on the analyses and experiments of large-scale real data, which remains as a future work.

\section{Conclusion And future Work}
We first proposed four strategies to collect schema data from the Web, getting a great deal of XML schemas, that leads to more accurate results than previous work. Then, we conducted an extensive study to investigate the practical usage of DREs, based on the large corpus of real data. The observations indicate that DREs are commonly used in practice and it is necessary to further study new practical subclasses of DREs. Then we gave some discussions and applications of our experiments, including detailed discussions on subclasses, a DRE data set build from the original data set, and SchemaRank to analyze the importance of XML Schemas. On the DRE data set, we use normalized DREs to compact and enhance it and use pattern generalization to expand it, which make us can regard the original alphabets covered by the data set.

Future work includes: 1) obtaining various sorts of real data; 2) more applications for our original data set or DRE data set; 3) further studying more new practical subclasses of DREs, in especial non-SORE subclasses of DREs, which is currently absent; 4) researching the reference relationship between Schemas and the definition and constraint relationship between XML files and Schemas based on our collected data; 5) studying the quality of XML and Schema on the Web, starting with the data that has been obtained.

\bibliographystyle{ACM-Reference-Format}
\bibliography{sample-bibliography}

\end{document}